\documentclass[twocolumn,showpacs,pra,aps,superscriptaddress]{revtex4}
\usepackage{ulem}
\usepackage{array}
\usepackage{amsmath}
\usepackage{graphicx}
\usepackage{dsfont}
\usepackage{color}
\usepackage{subfigure}
\def\bra#1{\langle #1|}
\def\ket#1{|#1 \rangle}

\begin{document}

\title{Testing quantum nonlocality by generalized quasiprobability functions}

\author{Seung-Woo Lee}
\email{swleego@gmail.com} \affiliation{Clarendon Laboratory,
University of Oxford, Parks Road, Oxford OX1 3PU, United Kingdom}

\author{Hyunseok Jeong}

\affiliation{Center for Subwavelength Optics and Department of
Physics and Astronomy, Seoul National University, Seoul, 151-742,
Korea}

\author{Dieter Jaksch}

\affiliation{Clarendon Laboratory, University of Oxford, Parks Road,
Oxford OX1 3PU, United Kingdom} \affiliation{Center for Quantum
Technologies, National University of Singapore, Singapore 117543,
Singapore}

\date{\today}

\begin{abstract}
We derive a Bell inequality based on a generalized quasiprobability
function which is parameterized by one non-positive real value. Two
types of known Bell inequalities formulated in terms of the Wigner-
and Q-functions are included as limiting cases. We investigate
violations of our Bell inequalities for single photon entangled
states and two-mode squeezed vacuum states when varying the detector
efficiency. We show that the Bell inequality for the Q-function
allows the lowest detection efficiency for violations of local
realism.
\end{abstract}


\maketitle

\section{Introduction}

Ever since the famous arguments of Einstein-Podolski-Rosen (EPR)
\cite{EPR35}, quantum nonlocality has been a central issue for
understanding the conceptual foundations of quantum mechanics.
Quantum nonlocality can be demonstrated by the violation of Bell
inequalities (BIs) \cite{Bell64} which are obeyed by local-realistic
(LR) theories. Realizations of BI tests are thus of great importance
in testing the validity of quantum theories against LR theories. In
addition, BI tests play a practical role in the detection of
entanglement which is one of the main resources for quantum
information processing. Bell inequality tests for 2-dimensional
systems have already been realized \cite{experiments}, while BI
tests in higher-dimensional and continuous variable systems remain
an active area of research \cite{Gisin07,Braunstein05}.

Phase space representations are a convenient tool for the analysis
of continuous variable states as they provide insights into the
boundaries between quantum and classical physics. Any quantum state
$\hat{\rho}$ can be fully characterized by the quasiprobability
function defined in phase space \cite{Cahill69}. In contrast to the
probability functions in classical phase space the quasiprobability
function is not always positive. For example, the Wigner-function of
the single photon state has negative values in certain regions of
phase space \cite{Lvovsky2001}. Since the negativity of the
quasiprobability function inevitably reflects a non-classical
feature of quantum states, the relation between negativity of
quasi-probabilities and quantum nonlocality has been investigated
\cite{Bell87,Banaszek99}. Bell argued \cite{Bell87} that the
original EPR state will not exhibit nonlocality since its
Wigner-function is positive everywhere and hence serves as a
classical probability distribution for hidden variables. On the
other hand, Banaszek and W\'{o}dkiewicz (BW) showed how to
demonstrate quantum nonlocality using the Q- and Wigner-functions
\cite{Banaszek99}. They suggested two distinct types of BIs, one of
which is formulated via the Q-function and referred to in this paper
as the BW-Q inequality while the other is formulated using the
Wigner-function and is referred to as the BW-W inequality.
Remarkably, the BW-W inequality was shown to be violated by the EPR
state \cite{Banaszek99}. This indicates that there is no direct
relation between the negativity of the Wigner-function and
nonlocality.

Quasiprobability functions can be parameterized by one real
parameter $s$ \cite{Cahill69,Moya93}
\begin{eqnarray}
\label{eq:sQP} W(\alpha; s)=\frac{2}{\pi(1-s)}\mathrm{Tr}[\hat{\rho}
\hat{\Pi}(\alpha; s)],
\end{eqnarray}
where $\hat{\Pi}(\alpha; s) = \sum^{\infty}_{n=0}((s+1)/(s-1))^n
\ket{\alpha,n}\bra{\alpha,n}$, and $\ket{\alpha,n}$ is the number
state displaced by the complex variable $\alpha$ in phase space. It
is produced by applying the Glauber displacement operator $\hat
D(\alpha)$ to the number state $|n\rangle$. We call $W(\alpha; s)$
the $s$-parameterized quasiprobability function which becomes the
P-function, the Wigner-function, and the Q-function when setting
$s=1, 0, -1$ \cite{Moya93}, respectively. For non-positive $s$ the
function $W(\alpha; s)$ can be written as a convolution of the
Wigner-function and a Gaussian weight
\begin{eqnarray}
 W(\alpha; s) = \frac{2}{\pi |s|}\int d^2\beta
~W(\beta)\exp\left(-{\frac{2|\alpha-\beta|^2}{|s|}}\right).
\end{eqnarray}
This can be identified with a smoothed Wigner-function affected by
noise which is modeled by Gaussian smoothing
\cite{Leonhardt93,Banaszek96,Banaszek02}. Therefore decreasing $s$
reduces the negativity of the Wigner function and is thus often
considered to be a loss of quantumness. For example, the Q-function
($s=-1$), which is positive everywhere in phase space, can be
identified with the Wigner function smoothed over the area of
measurement uncertainty.

The purpose of this paper is to propose a method for testing quantum
nonlocality using the $s$-parameterized quasiprobability function.
We will firstly formulate a generalized BI in terms of the
$s$-parameterized quasiprobability function in
Sec.~\ref{section:GBI}. This will lead us to a {\em
$s$-parameterized Bell inequality} which includes the BW-Q and the
BW-W inequalities as limiting cases. We will then present a
measurement scheme to test BIs using imperfect detectors in
Sec.~\ref{section:TQN}. The measured Bell expectation value can be
written as a function of the parameter $s$ and the overall detector
efficiency $\eta$. In Sec.~\ref{section:VSP} violations of BIs will
be demonstrated for single-photon entangled states and in
Sec.~\ref{section:VTMSS} for two-mode squeezed vacuum states. We
find the range of $s$ and $\eta$ which allows observing non-local
properties of these two types of states. We will show that the test
involving the Q-function permits the lowest detector efficiency for
observing violations of local realism. We also find that the degree
of violation is irrespective of the negativity of the
quasiprobability function. Finally, in Sec.~\ref{section:Remarks},
we discuss the characteristics and applications of the
$s$-parameterized BI.

\section{Generalized Bell inequalities of quasiprobability functions}
\label{section:GBI}

We begin by formulating a generalised BI in terms of
quasiprobability functions. Suppose that two spatially separated
parties, Alice and Bob, independently choose one of two observables,
denoted by $\hat{A}_1$, $\hat{A}_2$ and $\hat{B}_1$, $\hat{B}_2$
respectively. No restriction is placed on the number of possible
measurement outcomes (which may be infinite). We assume that the
measurement operators of the local observables $\hat{A}_1,
\hat{A}_2, \hat{B}_1, \hat{B}_2$ can be written as
\begin{eqnarray}
 \label{eq:LOOp}
 \nonumber
  \hat{A}_a =\hat{O}(\alpha_a; s),~~~~\hat{B}_a =\hat{O}(\beta_b;
  s),~~~\mathrm{for}~a,b=1,2
\end{eqnarray}
using a Hermitian operator
\begin{eqnarray}
 \label{eq:ObOp}
  \hat{O}(\alpha;s) =
  \begin{cases}
(1-s)
\hat{\Pi}(\alpha;s)+s\openone & \text{if $-1 < s \leq 0$},\\
2\hat{\Pi}(\alpha;s)-\openone & \text{if $ s \leq -1$},
\end{cases}
\end{eqnarray}
parameterized by a real non-positive number $s$ and an arbitrary
complex variable $\alpha$. Here $\openone$ is the identity operator.
The possible measurement outcomes of $\hat{O}(\alpha;s)$ are given
by its eigenvalues,
\begin{eqnarray}
  \label{eq:EVal}
\lambda_n =
    \begin{cases}
(1-s)(\frac{s+1}{s-1})^n+s & \text{if $-1 < s \leq 0$},\\
2(\frac{s+1}{s-1})^n-1 & \text{if $ s \leq -1$},
\end{cases}
\end{eqnarray}
and their eigenvectors are the displaced number states. The maximum
and minimum measurement outcomes of $\hat{O}(\alpha;s)$ for any
non-positive $s$ are $\lambda_{\mathrm{max}} = 1$ and
$\lambda_{\mathrm{min}} = -1$, respectively. For $s=0$ we have
$\hat{O}(\alpha;0)=\hat{\Pi}(\alpha; 0)=\sum^{\infty}_{n=0}(-1)^n
\ket{\alpha,n}\bra{\alpha,n}$, the displaced parity operator, while
for $s=-1$ we find that
$\hat{O}(\alpha;-1)=2\ket{\alpha}\bra{\alpha}-\openone$ projects
onto the coherent states.

A Bell operator can be constructed using the measurement operators
$\hat{A}_a$, $\hat{B}_b$ by way of a construction similar to the
CHSH combination
\begin{eqnarray}
 \label{eq:BOp}
  {\cal \hat{B}}= \hat{C}_{1,1}+\hat{C}_{1,2}+\hat{C}_{2,1}-\hat{C}_{2,2},
\end{eqnarray}
where $\hat{C}_{a,b} =\hat{A}_a\otimes\hat{B}_b$ is the correlation
operator. Since the expectation values of the local observables are
bounded by $|\langle \hat{A}_a \rangle|<1$ and $|\langle \hat{B}_b
\rangle| \leq 1$ for any non-positive $s$, the expectation value of
the Bell operator defined in Eq.~(\ref{eq:BOp}) is bounded by
$|\langle \hat{{\cal B}}\rangle| \equiv |{\cal B}| \leq 2$ in LR
theories. Note that the expectation value of $\hat{\Pi}(\alpha; s)$
for a given density operator $\hat{\rho}$ is proportional to the
$s$-parameterized quasiprobability function \cite{Cahill69,Moya93}
\begin{eqnarray}
  \label{eq:Top}
  \nonumber
  W(\alpha; s) &=&\frac{2}{\pi(1-s)}\mathrm{Tr}[\hat{\rho} \hat{\Pi}(\alpha; s)]\\
  &=&\frac{2}{\pi(1-s)}\sum^{\infty}_{n=0}\left(\frac{s+1}{s-1}\right)^n
  \bra{\alpha,n}\hat{\rho} \ket{\alpha,n},
\end{eqnarray}
from which both the Wigner function and the Q-function can be
recovered by setting $s=0$ and $s=-1$, respectively. We do not
consider the case $s>0$ when the eigenvalues of $\hat{\Pi}(\alpha;
s)$ are not bounded. We thus obtain the following generalised BI
\begin{eqnarray}
\label{eq:BFinQP}
\nonumber
  |{\cal B}|_{\{-1<s\leq0\}}&=& \biggl|
  \frac{\pi^2(1-s)^4}{4}[W(\alpha_1,\beta_1;s)+W(\alpha_1,\beta_2;s)\\
\nonumber
  &+&W(\alpha_2,\beta_1;s)-W(\alpha_2,\beta_2;s)]+\pi s(1-s)^2\\
\nonumber
  &\times&[W(\alpha_1;s)+W(\beta_1;s)]+2s^2 \biggr|
  ~\leq 2,\\\\
\nonumber
  |{\cal B}|_{\{s\leq-1\}}&=& |
  \pi^2(1-s)^2[W(\alpha_1,\beta_1;s)+W(\alpha_1,\beta_2;s)\\
\nonumber
  &+&W(\alpha_2,\beta_1;s)-W(\alpha_2,\beta_2;s)]-2\pi(1-s)\\
\nonumber
  &\times&[W(\alpha_1;s)+W(\beta_1;s)]+2| ~\leq 2,
\end{eqnarray}
where $W(\alpha,\beta; s) =(4/\pi^2(1-s)^2)\mathrm{Tr}[\hat{\rho}
\hat{\Pi}(\alpha; s)
  \otimes\hat{\Pi}(\beta; s)]$ is the two-mode $s$-parameterized
quasiprobability functions, and $W(\alpha;s)$ and $W(\beta;s)$ are
its marginal distributions. We call Eq.~(\ref{eq:BFinQP}) the {\em
$s$-parameterized Bell inequality} for quasiprobability functions.
This BI is equivalent to the BW-W inequality when $s=0$ which has
the form of the standard CHSH inequality \cite{CHSH69}, and the BW-Q
inequality when $s=-1$ in the form of the BI proposed by Clauser and
Horn (CH) \cite{CH74}. In these cases the corresponding generalized
quasiprobability function reduces to the Wigner-function
$W(\alpha,\beta)=W(\alpha,\beta;0)$ and the Q-function
$Q(\alpha,\beta)=W(\alpha,\beta;-1)$, respectively
\cite{Banaszek99}.

\section{Testing Quantum nonlocality}
\label{section:TQN}

\begin{figure}
\begin{center}
\includegraphics[width=0.35\textwidth]{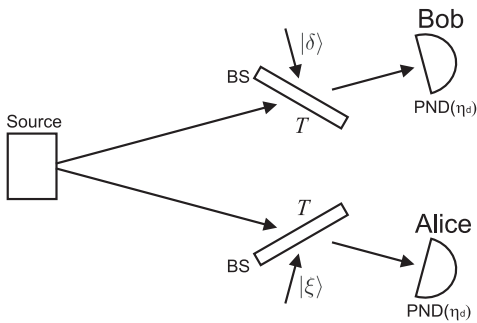}
\caption{The optical setup for the BI test. Each local measurement
is carried out after mixing the incoming field with a coherent state
(denoted by $\ket{\xi}$ for Alice and $\ket{\delta}$ for Bob) in a
beam splitter (BS) of high transmissivity $T$. The photon number
detectors (PNDs) have efficiency $\eta_d$.}\label{fig1}
\end{center}
\end{figure}

In this section we present a scheme to test quantum nonlocality
using the $s$-parameterized BIs. For a valid quantum nonlocality
test the measured quantities should satisfy the LR conditions which
are assumed when deriving BIs. Thus we here employ the direct
measurement scheme of quasiprobability functions using photon number
detectors proposed in \cite{Banaszek96}.

A pair of entangled states generated from a source of correlated
photons is distributed between Alice and Bob, each of whom make a
local measurement by way of an unbalanced homodyne detection (see
Fig.~\ref{fig1}). Each local measurement is carried out using a
photon number detector with quantum efficiency $\eta_d$ preceded by
a beam splitter with transmissivity $T$. Coherent fields $\ket{\xi}$
and $\ket{\delta}$ enter through the other input ports of each beam
splitter. For high transmissivity $T\rightarrow 1$ and strong
coherent fields $\xi$,$\delta \rightarrow \infty$, the beam
splitters of Alice and Bob can be described by the displacement
operators $\hat{D}(\alpha)$ and $\hat{D}(\beta)$ respectively, where
$\alpha = \xi\sqrt{(1-T)/T}$ and $\beta = \delta\sqrt{(1-T)/T}$
\cite{Banaszek96}. Measurements $((s+1)/(s-1))^{\hat{n}}$ with
$\hat{n}=\sum_{n} n\ket{n}\bra{n}$ the photon number operator are
performed on the outgoing modes using perfect photon number
detectors. Then the expectation value directly yields the value of
the $s$-parameterized quasiprobability function at the point in
phase space specified by the complex variables $\alpha$ and $\beta$.
For example, the Wigner function can be obtained by the parity
measurements $(-1)^{\hat{n}}$ ($s=0$) and the Q-function by on-off
(i.e. photon presence or absence) measurements ($s=-1$).

Let us now consider the effects of the detector efficiencies $\eta$.
If the true photon number distribution is given by $P(n)$, then the
measured distribution can be written as a function of the overall
detection efficiency $\eta=\eta_d T$ as
$P_{\eta}(m)=\sum^{\infty}_{n=m}P(n)\binom{n}{m}(1-\eta)^{n-m}\eta^m$
\cite{pndist}. For $\alpha=0$ the measured quasiprobability function
is
\begin{eqnarray}
 \label{eq:Wexp}
 \nonumber
  W_{\eta}(0;s)&=&\frac{2}{\pi(1-s)}\sum^{\infty}_{m=0}\left(\frac{s+1}{s-1}\right)^mP_{\eta}(m)\\
 \nonumber
  &=&\frac{2}{\pi(1-s)}\sum^{\infty}_{n=0}\left(1-\eta+\eta\frac{s+1}{s-1}\right)^n P(n)\\
  &=& \frac{W\left(0;-\frac{1-s-\eta}{\eta} \right)}{\eta} \equiv  \frac{W\left(0;s' \right)}{\eta}.
\end{eqnarray}
The $s$-parameterized quasiprobability function measured by a
detector with efficiency $\eta$ can therefore be identified with the
quasiprobability function with parameter $s'=-(1-s-\eta)/\eta$.
Other sources of noise (e.g.~dark counts and mode mismatch) could be
included into this approach but are neglected here for simplicity.

Finally, the expectation value of observable (\ref{eq:ObOp}) is
given as
\begin{equation}
 \label{eq:expecob}
 \langle \hat{O}(\alpha;s)\rangle_{\eta} =
  \begin{cases}
\frac{\pi(1-s)^2}{2\eta}W(\alpha;s')+s
& \text{if $-1 < s \leq 0$},\\
\frac{\pi(1-s)}{\eta}W(\alpha;s')-1& \text{if $ s \leq -1$},
\end{cases}
\end{equation}
where $\langle \cdot \rangle_{\eta}$ represents the expectation
value obtained by measurement with efficiency $\eta$. Note that
(\ref{eq:expecob}) is the statistical average of directly measured
data without postselection. The expectation value of the Bell
operator (\ref{eq:BOp}) written as a function of $s$ and $\eta$ is
given by
\begin{widetext}
\begin{eqnarray}
\label{eq:BFinSW} \nonumber
  \langle \hat{{\cal B}}_{\{-1<s\leq0\}} \rangle_{\eta} &=&
  \frac{\pi^2(1-s)^4}{4\eta^2}\biggl[W(\alpha_1,\beta_1;-\frac{1-s-\eta}{\eta})+W(\alpha_1,\beta_2;-\frac{1-s-\eta}{\eta})
  +W(\alpha_2,\beta_1;-\frac{1-s-\eta}{\eta})\\
  \nonumber
  &&~~~~~~~~-W(\alpha_2,\beta_2;-\frac{1-s-\eta}{\eta})\biggr]+\frac{\pi s(1-s)^2}{\eta}\biggl[W(\alpha_1;-\frac{1-s-\eta}{\eta})
  +W(\beta_1;-\frac{1-s-\eta}{\eta})\biggr]+2s^2,\\
  \langle \hat{{\cal B}}_{\{s\leq-1\}} \rangle_{\eta} &=&\frac{\pi^2(1-s)^2}{\eta^2}\biggl[W(\alpha_1,\beta_1;-\frac{1-s-\eta}{\eta})
  +W(\alpha_1,\beta_2;-\frac{1-s-\eta}{\eta})+W(\alpha_2,\beta_1;-\frac{1-s-\eta}{\eta})\\
  \nonumber
  &&~~~~~~~~-W(\alpha_2,\beta_2;-\frac{1-s-\eta}{\eta})\biggr]-\frac{2\pi(1-s)}{\eta}\biggl[W(\alpha_1;-\frac{1-s-\eta}{\eta})
  +W(\beta_1;-\frac{1-s-\eta}{\eta})\biggr]+2.
\end{eqnarray}
\end{widetext}
Note that the Bell expectation values in Eq.~(\ref{eq:BFinSW}) for
$s=0$ and $s=-1$ give the same results as tests of the BW-W and BW-Q
inequalities, respectively.

\section{Violation by single photon entangled states}
\label{section:VSP}

\begin{figure}
\begin{center}
\includegraphics[width=0.45\textwidth]{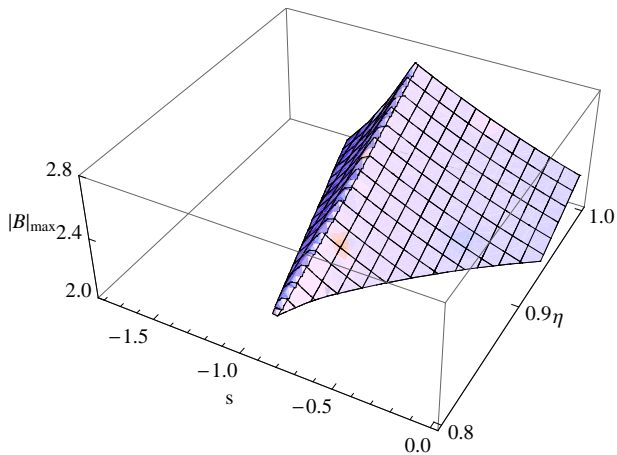}
\caption{Maximum Bell expectation value $|{\cal B}|=| \langle
\hat{{\cal B}} \rangle |$ for the single photon entangled state.
Only the range of parameters $s$ and detector efficiencies $\eta$
with $|{\cal B}| > 2$ is shown.}\label{fig2}
\end{center}
\end{figure}

We investigate violations of the $s$-parameterized BI
(\ref{eq:BFinQP}) for the single photon entangled state
\cite{singlephotone}
\begin{eqnarray}
\ket{\Psi}=\frac{1}{\sqrt{2}}(\ket{0,1}+\ket{1,0}),
\end{eqnarray}
where $\ket{n,m}$ is the state with $n$ photons in Alice's mode and
$m$ photons in Bob's mode. This state is created by a single photon
incident on a 50:50 beam splitter. Its two-mode $s$-parameterized
quasiprobability function is given by
\begin{eqnarray}
\nonumber \label{eq:QPofSPES}
W_{\Psi}(\alpha,\beta;s)&=&\frac{4}{\pi^2(1-s)^2}\left(-\frac{1+s}{1-s}+\frac{2}{(1-s)^2}|\alpha+\beta|^2\right)\\
&&~~~~~~~~\times\exp\left[-\frac{2(|\alpha|^2+|\beta|^2)}{1-s}\right],
\end{eqnarray}
and its marginal single-mode distribution is
\begin{eqnarray}
W_{\Psi}(\alpha;s)=(1/\pi)(2-2\eta+4\eta^2|\alpha|^2)
\exp[-2\eta|\alpha|^2].
\end{eqnarray}
Note that for $0\geq s>-1$ Eq.~(\ref{eq:QPofSPES}) has negative
values in certain regions of phase space but for $s=-1$ it becomes
the Q-function $W_{\Psi}(\alpha,\beta;-1)\geq 0$.

The maximum expectation values $|{\cal B}|_{\mathrm{max}}=| \langle
\hat{{\cal B}} \rangle |_{\mathrm{max}}$ are obtained for properly
chosen $\alpha_1$, $\alpha_2$, $\beta_1$, $\beta_2$.
Figure~\ref{fig2} shows the range of parameters $s$ and detector
efficiencies $\eta$ for which the BI is violated, $|{\cal
B}|_{\mathrm{max}} > 2$. Interestingly, the degree of violation is
not directly related to the negativity of the quasiprobability
functions. The test of the BI using the Q-function ($s=-1$) yields
strong violations and is most robust to detector inefficiencies.
This is because the observable (\ref{eq:ObOp}) becomes dichotomized
at $s =-1$ corresponding to detection of none vs.~some photons. For
a given $s$, the amount of violation decreases with decreasing
$\eta$. The minimum value of $\eta$ indicates the required detector
efficiency for a successful nonlocality test \cite{Larsson98}. For
example, the minimum bound is about $83\%$ for the Q-function
($s=-1$). We also find the minimum parameter $s$ which allows
demonstrating quantum nonlocality for a given detector efficiency.
For example, for a perfect detector ($\eta=1$), the corresponding BI
is violated when $s \gtrsim -1.43$.

\section{Violation by two-mode squeezed states}
\label{section:VTMSS}

\begin{figure*}
\begin{center}
\includegraphics[width=\linewidth]{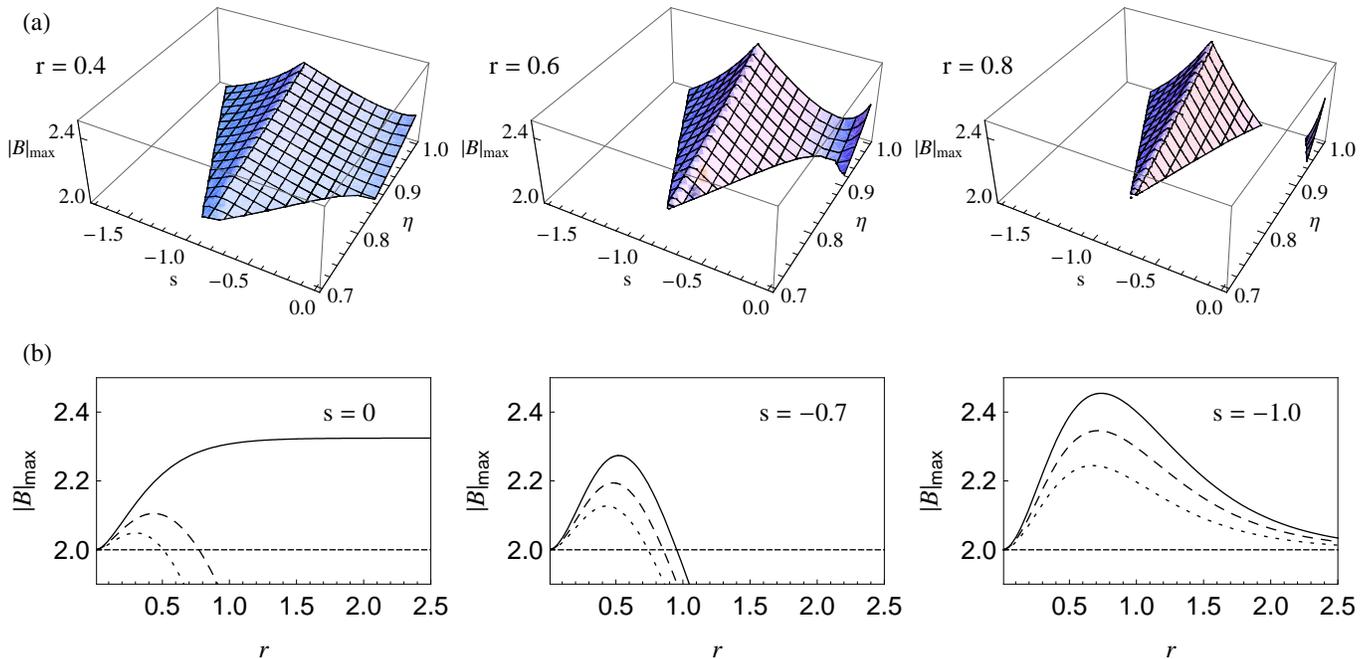}
\caption{Demonstration of quantum nonlocality for TMSSs. (a) Maximum
Bell values are shown for different squeezing $r$ in the range of
$s$ and $\eta$ where the BI is violated. (b) Violation of the BI as
a function of the squeezing $r$ for different $s$ and $\eta=1$
(solid line), $\eta=0.95$ (dashed line), and $\eta=0.9$ (dotted
line).}\label{fig3}
\end{center}
\end{figure*}

We consider the two-mode squeezed vacuum states (TMSSs), i.e. a
continuous variable entangled state written as
\begin{eqnarray}
\label{eq:TMSSs}
\ket{\mathrm{TMSS}}=\mathrm{sech}~r\sum_{n=0}^{\infty}\tanh^{n}{r}\ket{n,n},
\end{eqnarray}
where $r>0$ is the squeezing parameter. It can be realized for
instance by non-degenerate optical parametric amplifiers
\cite{Reid88}. In the infinite squeezing limit $r\rightarrow
\infty$, the TMSS becomes the normalized EPR state which is the
maximally entangled state associated with position and momentum
\cite{Banaszek99}.

For a non-positive $s$ the quasiprobability function of the TMSS is
given by
\begin{eqnarray}
 \label{eq:QpfTMSS}
\nonumber
W_{\mathrm{TMSS}}(\alpha,\beta;s)&=&\frac{4}{\pi^2R(s)}\exp\biggl(-\frac{2}{R(s)}
  \{S(s)(|\alpha|^2+|\beta|^2)\\
  &&-\sinh{2r}(\alpha\beta+\alpha^*\beta^*)\}\biggr),
\end{eqnarray}
and its marginal single-mode distribution is
\begin{eqnarray}
 \label{eq:sQpfTMSS}
  W_{\mathrm{TMSS}}(\alpha;s)=\frac{2}{\pi S(s)}\exp\biggl(-\frac{2|\alpha|^2}{S(s)}\biggr),
\end{eqnarray}
where $R(s)=s^2-2s\cosh{2r}+1$ and $S(s)=\cosh{2r}-s$. Note that
these are positive everywhere in phase space. In Fig.~\ref{fig3}(a)
violations of the $s$-parameterized BI are shown for TMSSs. The test
using the Q-function ($s=-1$) is most robust with respect to
detector inefficiencies. The amount of violation shows different
tendencies depending on the squeezing parameter $r$. In the case of
low squeezing rates, i.e.\ when the amplitudes of small-n number
states are dominant, the violation is maximal if we choose the
Q-function ($s=-1$) as shown in Fig.~\ref{fig3}(b). This implies
that the dominant contribution to the violation comes from
correlations between the vacuum and photons being present. For
larger squeezing rates $r \gtrsim 1.2$, the violation reaches a
maximal value ${\cal B} \approx 2.32$ when we test the
Wigner-function ($s=0$) \cite{Jeong03}. This indicates that the
parity measurements are effective for verifying higher-order number
correlations. However, the parity measurements require very high
detector efficiency as shown in Fig.~\ref{fig3}(b). The range of $s$
within which one can demonstrate nonlocality becomes narrower around
$s=0$ and $s=-1$ with increasing squeezing rate $r$. This is because
the observable (\ref{eq:ObOp}) is dichotomized at $s=0$ and $s=-1$.

\section{Discussion and Conclusions}
\label{section:Remarks}

We demonstrated that quantum nonlocality has no direct relation to
the negativity of $s$-parameterized quasiprobability functions. In
fact the Q-function ($s=-1$) which never becomes negative can still
be used to verify non-local properties as we showed in
Fig.~\ref{fig2} and yields strong violations of the corresponding
BI. This implies that the quantum properties of nonlocality and
negativity of the quasiprobability functions should be considered
distinct features of quantum mechanics. Furthermore we showed that
the Q-function test allows the lowest detector efficiency for
demonstrating quantum nonlocality. For example, it requires only
$\eta\approx 83$\% for a single photon entangled state and
$\eta\approx 75$\% for TMSSs with $r=0.4$ to detect nonlocality.
This indicates that two-mode correlations between vacuum and many
photons can be more robust to detector inefficiencies than
correlations between vacuum and a single photon.

The parameter $s$ determines the characteristics of the detected
non-local correlations. For example, if we choose $s=-1$ the
violation of the BI exhibits only correlations between vacuum and
photons. In order to test higher-order photon number correlations we
need to increase $s$ to zero, so that the factor $((s+1)/(s-1))^n$
multiplied to the photon number probability increases in
Eq.~(\ref{eq:Top}). Although parity measurements ($s=0$) allow to
detect higher-order correlations effectively, they also require very
high detector efficiencies as shown in Fig.~\ref{fig3}. If we
properly choose a certain parameter $-1<s<0$, e.g. $s=-0.7$, we can
detect higher-order correlations with a lower detector efficiency
than that required for testing the BI using the Wigner function.
However, we note that the violation of the BI with $s=-0.7$
disappears with increasing squeezing rate as shown in
Fig.~\ref{fig3}(b); this restricts the possible applications to
schemes using light that contains only a few photons.

Let us finally discuss whether we can regard decoherence effects as
changes to $s$. Interactions with the environment and detection
noise tend to smoothen quasiprobability functions. For example, when
solving the Fokker-Planck equation for the evolution of the
Wigner-function of a system interacting with a thermal environment
one obtains \cite{Jeong00}
\begin{eqnarray}
W(\alpha, \tau)=\frac{1}{t(\tau)^2} \int d^2\beta
W^{\mathrm{th}}(\beta)W\biggl(\frac{\alpha-r(\tau)\beta}{t(\tau)},\tau=0
\biggr).
\end{eqnarray}
Here the parameters $r(\tau)=\sqrt{1-e^{-\gamma\tau}}$ and
$t(\tau)=\sqrt{e^{-\gamma\tau}}$ are given in terms of the energy
decay rate $\gamma$, and
\begin{eqnarray}
W^{\mathrm{th}}(\beta)=\frac{2}{\pi(1+2\bar{n})}\exp{\biggl(-\frac{2|\beta|^2}{1+2\bar{n}}\biggr)}
\end{eqnarray}
is the Wigner function for the thermal state of average thermal
photon number $\bar{n}$. The effect of the thermal environment is
then identified with temporal changes of the parameter
\begin{eqnarray}
s(\tau)\sim
-\frac{r(\tau)^2}{t(\tau)^2}(1+2\bar{n})=(1-e^{\gamma\tau})(1+2\bar{n}).
\end{eqnarray}
Therefore one might be tempted to consider an environment in a
thermal state as giving rise to a temporal change in $s$ in
Eq.~(\ref{eq:ObOp}). However this idea is not applicable to tests of
quantum nonlocality. The $s$-parameterized BI is derived for
observables (\ref{eq:ObOp}) which contain $s$ as a deterministic
value of LR theories. Thus the local-realistic bound is no longer
valid when dynamical observables are considered (even though they
give the same statistical average). However, this idea might be
useful for witnessing entanglement \cite{Lee09}.

In summary, we have formulated a BI in terms of the generalized
quasiprobability function. This BI is parameterized by a
non-positive value $s$ and includes previously proposed BIs such as
the BW-W ($s=0$) and the BW-Q ($s=-1$) inequalities
\cite{Banaszek99}. We employed a direct measurement scheme for
quasiprobability functions \cite{Banaszek96} to test quantum
nonlocality. The violation of BIs was demonstrated for two types of
entangled states, single photon entangled and two-mode squeezed
vacuum states. We found the range of $s$ and $\eta$ which allow the
observation of quantum non-local properties. We discussed the types
of correlations and their robustness to detection inefficiencies for
different values of $s$. We also demonstrated that the negativity of
the quasiprobability function is not directly related to the
violation of BIs. The realization of $s$-parameterized BI tests is
expected along with the progress of photon detection technologies
\cite{Divochiy08} in the near future. Our investigations can readily
be extended to other types of states like photon subtracted gaussian
states \cite{Wenger04,Jeong08}, or optical Schr\"{o}dinger cat
states \cite{Ourjoumtsev07}.

\acknowledgments We thank E. Knill, B. J. Smith, L. Heaney for
valuable comments. This work was supported by the UK EPSRC through
projects QIPIRC (GR/S82176/01), EuroQUAM (EP/E041612/1), the World
Class University (WCU) program and the KOSEF grant funded by the
Korea government(MEST) (R11-2008-095-01000-0).

\end{document}